\documentclass[aps,twocolumn,
showpacs,pra]{revtex4}
\usepackage{amsmath,latexsym,amssymb}
\usepackage{amsmath}
\usepackage{latexsym}
\usepackage{amssymb}

\begin{document}

\title{Multipartite omnidirectional generalized Bell inequality}

\author{Koji Nagata}
\affiliation{Department of Physics, Korea Advanced Institute of Science and Technology, Daejeon 305-701, Korea}

\date{\today}

\begin{abstract}
We derive a multipartite generalized Bell inequality 
which involves the entire range of settings for each of the local observers.
Especially, it is applied to show non-local behavior
of a six-qubit mixture of Greenberger-Horne-Zeilinger correlations
stronger than previous Bell inequalities.
For certain noise admixture to the correlations
an explicit local realistic model exists
in the case of a standard Bell experiment.
Bell experiments with many local settings
reveal the non-locality of the state.
It turns out that the new inequality
is more stringent than many other Bell inequalities 
in the specific quantum state.
\end{abstract}

\maketitle

\section{introduction}

Non-locality in quantum physics means the possibility of
distributing correlations that cannot be due to previously shared
randomness, without signaling \cite{bib:Peres3,bib:Redhead}.
Certain quantum predictions violate Bell inequalities \cite{bib:Bell}, 
which form necessary conditions for local realistic models for 
the results of suitable measurements.
Thus, certain measurement outcome in quantum predictions cannot admit local realistic theories.

In many cases one can build a local realistic model
for the observed data.
However, many such models are artificial
and can be disproved
if some principles
of physics are taken into account.
An example of such a principle
is rotational invariance of correlation function ---
the fact that the value of correlation function does not depend 
on the orientation of reference frames.
Taking this additional requirement into account
rules out local realistic models
even in situations
in which standard Bell inequalities allow for an explicit
construction of such models \cite{Nagata}.

Here, we derive a generalized Bell inequality for $N$ qubits
which involves the entire range of settings 
for each of the local measuring apparatuses.
The inequality forms a necessary condition
for the existence of a local realistic model
which predicts rotationally invariant correlations.
Although the inequality involves the
entire range of settings
it can be experimentally tested
using three orthogonal local measurement settings.
This is a direct consequence
of the assumed form 
of rotationally invariant correlations.

Next, we consider a mixture of 
Greenberger-Horne-Zeilinger (GHZ) states \cite{bib:GHZ}
written in three orthogonal directions.
A white noise is added to the mixture with some probability.
We take the minimal amount of noise admixture
for which one does not violate a Bell inequality
as a measure of the strength of the inequality.
It turns out that the new inequality
is more stringent than many other 
inequalities \cite{Nagata, Werner, WZ, Zukowski} in the specific quantum state.

\section{multipartite omnidirectional generalized Bell inequality}

Consider $N$ spin-$\frac{1}{2}$ particles,
each in a separate laboratory. 
Let us parameterize the local settings
of the $j$th observer with a unit vector $\vec n_j$ with $j=1,\ldots,N$. 
One can introduce the ``Bell'' correlation function, which is the average of the product of the local results
\begin{equation}
E(\vec n_1, \vec n_2,\ldots, \vec n_N) = 
\langle r_1(\vec n_1) r_2(\vec n_2)\cdots r_N(\vec n_N) \rangle_{\rm avg},
\end{equation}
where $r_j(\vec n_j)$ is the local result, $\pm 1$, which is obtained if the measurement direction is set at $\vec n_j$.
If correlation function 
admits a rotationally invariant tensor structure
familiar from quantum mechanics, we can introduce the following form:
\begin{equation}
E(\vec n_1, \vec n_2,\ldots, \vec n_N) 
= \hat T \cdot (\vec n_1 \otimes \vec n_2 \otimes \cdots \otimes \vec n_N),
\label{et}
\end{equation}
where $\otimes$ denotes the tensor product, 
$\cdot$ the scalar product in ${\rm R}^{3N}$ and 
$\hat T$ is the correlation tensor 
the elements of which are
given by
\begin{equation}
T_{i_1...i_N} \equiv E(\vec x_{1}^{(i_1)},\vec x_{2}^{(i_2)},\ldots, \vec x_{N}^{(i_N)}),
\label{tensor}
\end{equation}
where $\vec x_{j}^{(i_j)}$ is a unit vector of the local coordinate system of the $j$th observer;
$i_j = 1,2,3$ gives the full set of orthogonal vectors defining the local Cartesian coordinates. 
The components of the correlation tensor are experimentally accessible 
by measuring the correlation function at the directions given by the bases vectors 
in which the tensor is written
\footnote{The same idea is behind quantum tomography.}.
Suppose one knows the values of all $3^N$ components of the correlation tensor, $T_{i_1...i_N}$.
Then, with the help of formula (\ref{et}) one can compute the value of the correlation function for all other possible sets of local settings.

We shall derive a necessary condition for the existence of a local realistic 
description of the rotationally invariant correlation function (\ref{et}).
A correlation function has a local realistic model
if it can be written as
\begin{eqnarray}
&&E_{LR}(\vec{n}_1,\vec{n}_2,\ldots,\vec{n}_N)=\nonumber\\
&&\int d\lambda \rho(\lambda)
I^{(1)}(\vec{n}_1,\lambda)I^{(2)}(\vec{n}_2,\lambda)\cdots
I^{(N)}(\vec{n}_N,\lambda),
\label{LHVcofun}
\end{eqnarray}
where $\lambda$ denotes a set of hidden variables, 
$\rho(\lambda)$ is their distribution, and $I^{(j)}(\vec{n}_j,\lambda)$ is 
the predetermined ``hidden'' result of 
the measurement of all the dichotomic observables 
parameterized by any direction of $\vec n_j$.

One can write the observable (unit) vector $\vec n_j$
in a spherical coordinate system:
\begin{equation}
\vec{n}_j(\theta_j, \phi_j) = \sin \theta_j\cos\phi_j \vec{x}_j^{(1)}
+\sin \theta_j\sin\phi_j \vec{x}_j^{(2)}
+\cos\theta_j \vec{x}_j^{(3)},
\label{vector}
\end{equation}
where $\vec x_j^{(1)}$, $\vec x_j^{(2)}$, and $\vec x_j^{(3)}$
are the Cartesian axes relative to which
spherical angles are measured.

We shall show that the scalar product 
of the local realistic correlation function, 
$E_{LR}$ given in (\ref{LHVcofun}),
with the rotationally invariant correlation function, $E$ given in (\ref{et}),
is bounded by a specific number dependent on $\hat{T}$.
We use decomposition (\ref{vector})
and introduce the usual measure
$d\Omega_j=\sin\theta_jd\theta_jd\phi_j$
for the system of the $j$th observer.
It will be proven that
\begin{eqnarray}
(E_{LR}, E) & = & \int\!\!d\Omega_1 \cdots
\int\!\!d\Omega_N
E_{LR}(\theta_1, \phi_1,\ldots,\theta_N, \phi_N)\nonumber\\
& \times &
E(\theta_1, \phi_1,\ldots,\theta_N, \phi_N)\leq (2\pi)^N T_{\max},
\label{Bell-Zineq}
\end{eqnarray}
where $T_{\max}$ is the maximal 
possible value of the correlation tensor component,
maximized over choices of all possible local settings:
\begin{equation}
T_{\max}=\max_{\theta_1, \phi_1,\ldots,\theta_N, \phi_N}
E(\theta_1, \phi_1,\ldots,\theta_N, \phi_N).
\label{TE}
\end{equation}

A necessary condition 
for the existence of a local realistic 
description of the rotationally invariant correlation function,
i.e., for $E_{LR}$ to be equal to $E$, 
is that the following scalar products are equal: $(E_{LR}, E) =(E, E)$.
If one finds $(E_{LR}, E)< (E, E)$,
then the rotationally invariant correlation 
function cannot be explained by any local realistic theory.
Note that, due to the integrations in (\ref{Bell-Zineq}),
we are looking for a model for the entire
range of settings.

In what follows, we derive the upper bound of (\ref{Bell-Zineq}).
Since the local realistic model is an average over $\lambda$, 
it is enough to find the upper bound of the following expression:
\begin{eqnarray}
&&\int\!\!d\Omega_1
\cdots
\int\!\!d\Omega_N 
I^{(1)}(\theta_1, \phi_1)\cdots
I^{(N)}(\theta_N, \phi_N) \nonumber\\
&& \times \sum_{i_1,i_2,\ldots,i_N=1,2,3}T_{i_1i_2...i_N}
c^{i_1}_{1}c^{i_2}_{2}\cdots c^{i_N}_{N},\label{integral}
\end{eqnarray}
where
\begin{eqnarray}
\vec c_j=(c^1_{j}, c^2_{j}, c^3_{j})=(\sin \theta_j\cos\phi_j,
\sin \theta_j\sin\phi_j,
\cos\theta_j), 
\end{eqnarray}
and
\begin{eqnarray}
T_{i_1i_2...i_N}=\hat{T} \cdot
(\vec{x}_1^{(i_1)}\otimes\vec{x}_2^{(i_2)}\otimes
\cdots\otimes\vec{x}_N^{(i_N)}),
\end{eqnarray}
compare with Eq.~(\ref{et}) and Eq.~(\ref{tensor}). 
Here, we use the abbreviation $I^{(j)}(\theta_j, \phi_j)$ for $I^{(j)}(\vec{n}_j(\theta_j, \phi_j), \lambda)$.

Let us analyze the structure of expression (\ref{integral}).
Notice that (\ref{integral}) is a sum, 
with coefficients given by $T_{i_1i_2...i_N}$, 
of products of the following integrals:
$
\int\!\!d\Omega_j 
I^{(j)}(\theta_j, \phi_j) \sin \theta_j\cos\phi_j,
\int\!\!d\Omega_j I^{(j)}(\theta_j, \phi_j) 
\sin \theta_j\sin\phi_j,
$
and
$
\int\!\!d\Omega_j 
I^{(j)}(\theta_j, \phi_j) \cos \theta_j.
$
These integrals are scalar products of
$I^{(j)}(\theta_j, \phi_j)$ with three orthogonal functions.
One has
$
\int\!\!d\Omega_j c^{i_k}_j c^{i_k'}_j=(4\pi/3)\delta_{i_k,i_k'}.
$
The normalized functions 
$\sqrt{3/4\pi}\sin\theta_j\cos\phi_j$, 
$\sqrt{3/4\pi}\sin\theta_j\sin\phi_j$, and
$\sqrt{3/4\pi}\cos\theta_j$ form a basis of a 
three-dimensional real functional space, which we shall call $S^{(3)}$ \cite{KZ}.
Using these three functions
one can write the projection of function $I^{(j)}(\theta_j, \phi_j)$
onto them as
\begin{eqnarray}
\int\!\!d\Omega_j 
I^{(j)}(\theta_j, \phi_j)
\sqrt{3/4\pi}\sin\theta_j\cos\phi_j
&=&\sin \beta_j\cos\gamma_j \Vert I^{(j){||}} \Vert,\nonumber\\ 
\int\!\!d\Omega_j 
I^{(j)}(\theta_j, \phi_j)
\sqrt{3/4\pi}\sin\theta_j\sin\phi_j
&=&\sin \beta_j\sin\gamma_j \Vert I^{(j){||}} \Vert,\nonumber \\
\int\!\!d\Omega_j 
I^{(j)}(\theta_j, \phi_j)
\sqrt{3/4\pi}\cos\theta_j
&=&\cos\beta_j \Vert I^{(j){||}} \Vert,
\end{eqnarray}
where $\Vert I^{(j){||}} \Vert$
is the length of the projection,
and $\beta_j$ and $\gamma_j$ are some angles.
Going back to expression (\ref{integral})
one has
\begin{eqnarray}
&& \left(\frac{4\pi}{3} \right)^{N/2}\prod_{j=1}^N \Vert I^{(j){||}} \Vert\nonumber\\
&& \times \sum_{i_1,i_2,\ldots,i_N=1,2,3}
T_{i_1i_2...i_N}
e^{i_1}_{1}e^{i_2}_{2}\cdots e^{i_N}_{N},
\label{EE}
\end{eqnarray}
with a normalized vector
\begin{eqnarray}
(e^{1}_j, e^{2}_j, e^{3}_j) = (\sin \beta_j\cos\gamma_j, \sin \beta_j\sin\gamma_j, \cos\beta_j).
\label{dcos}
\end{eqnarray}
Note that the sum in (\ref{EE}) 
over the components of this vector
is just $\hat T \cdot (\vec e_1 \otimes \vec e_2 \otimes \cdots \otimes \vec e_N)$,
i.e.,
it is a component of the tensor $\hat T$ 
in the local Cartesian coordinate systems
specified by the vectors $\vec e_j$.
If one knows all the values of $T_{i_1i_2...i_N}$, 
one can always find the maximal possible value of such a component, 
and it is equal to $T_{\max}$, of Eq.~(\ref{TE}).
Thus, 
\begin{equation}
\sum_{i_1,i_2,\ldots,i_N=1,2,3}
T_{i_1i_2...i_N}
e^{i_1}_{1}e^{i_2}_{2}\cdots e^{i_N}_{N} \le T_{\max}.
\end{equation}

It remains to show the upper bound on the norm $\Vert I^{(j){||}} \Vert$.
From the definition
the norm is given by a maximal 
possible value of the scalar product between
$I^{(j)}(\theta_j, \phi_j)$ and any normalized function belonging to $S^{(3)}$:
\begin{eqnarray}
\Vert I^{(j){||}} \Vert= \max_{|\vec d|=1}
\left[
\sqrt{\frac{3}{4\pi}}\int\!\!d\Omega_jI^{(j)}(\theta_j, \phi_j)
\sum_{k=1}^3 d_k c^k_j
\right],
\end{eqnarray}
where $\vec d = (d_1,d_2,d_3)$ and $|\vec d|=\sum_{k=1}^3 d_k^2=1$.
Since $|I^{(j)}(\theta_j, \phi_j)|=1$,
one has for the integral of the modulus
\begin{eqnarray}
\Vert I^{(j){||}} \Vert\leq \max_{|\vec d|=1}
\left[
\sqrt{\frac{3}{4\pi}}\int\!\!d\Omega_j
\left|
\vec d \cdot \vec c_j \right|
\right],\label{fack}
\end{eqnarray}
where the dot between three-dimensional vectors
denotes the usual scalar product in ${\rm R}^3$.
The values of this scalar product are then integrated (summed)
over all values of $\theta_j$ and $\phi_j$,
i.e., over vectors $\vec c_j$ on the whole sphere.
Since the measure is rotationally invariant 
the integral does not depend
on particular $\vec d$ and we choose it 
as a unit vector in direction $\vec z$.
For this choice
\begin{eqnarray}
\Vert I^{(j){||}} \Vert\leq \int\!\!d\Omega_j
\left|
\sqrt{\frac{3}{4\pi}}
\cos\theta_j \right|=2\pi\sqrt{\frac{3}{4\pi}}.
\end{eqnarray}
Finally $(E_{LR}, E) \leq (2\pi)^N T_{\max}$. $\square$

The relation (\ref{Bell-Zineq})
is a generalized $N$-qubit Bell inequality with the entire range of measurement settings.
Specific local hidden variable models, $E_{LR}$,
which rebuild rotationally invariant correlations, $E$,
satisfy it.
However, there exist rotationally invariant correlations
which cannot be modeled in a local realistic way.
Whenever the scalar product $(E, E)$
is bigger than the product $(E_{LR}, E)$
there can be no local realistic model for $E$.
Thus, we compute
\begin{eqnarray}
(E, E) & = &
\int\!\!d\Omega_1\cdots
\int\!\!d\Omega_N
\left(\sum_{i_1,\ldots,i_N=1}^3T_{i_1...i_N}
c^{i_1}_{1}\cdots c^{i_N}_{N}\right)^2 \nonumber\\
& = & (4\pi/3)^N \sum_{i_1,\ldots,i_N=1}^3T_{i_1...i_N}^2,
\label{EEvalue}
\end{eqnarray}
where we have used the orthogonality relation
$\int d \Omega_j ~ c_j^{i_k} c_j^{i_k'}  = (4\pi/3) \delta_{i_k,i_k'}$.
Finally, the necessary condition
for the existence of a local realistic model
of rotationally invariant correlations
which involve the entire range of settings reads
\begin{equation}
\max \sum_{i_1,i_2,\ldots,i_N=1,2,3}T_{i_1i_2...i_N}^2
\le \left( \frac{3}{2} \right)^N T_{\max},
\label{3D_NECC}
\end{equation}
where the maximization is taken over
all independent rotations of local coordinate systems
(or equivalently over all possible measurement directions).

\section{mixture of six-qubit GHZ state}

Now, we shall present the specific quantum state
for which the newly derived inequality
is better than the previous inequalities
described in Refs.~\cite{WZ, Werner, Zukowski, Nagata}.

Consider the following
six-qubit GHZ state
\begin{equation}
| \psi_3 \rangle = \frac{1}{\sqrt{2}} \Big( 
| z+ \rangle_1 \ldots | z+ \rangle_5 | z- \rangle_6 
+ | z- \rangle_1 \ldots | z- \rangle_5 | z+ \rangle_6 \Big),
\end{equation}
where $| z \pm \rangle_j$ is the eigenstate
of the local $\sigma_z$ operator
of the $j$th observer.
Note that the states of the last party are flipped with respect
to the states of the other parties.
We rotate the states of all individual
qubits by the angle $\alpha = 2 \pi / 3$
around the axis $\vec m = \frac{1}{\sqrt{3}}(1,1,1)$
on the Bloch sphere.
This rotation cyclically permutes the 
directions of the Cartesian coordinate system.
The unitary realizing this rotation
is given by \cite{NIELSEN_CHUANG}:
\begin{equation}
U = e^{- i \frac{\alpha}{2} \vec m \cdot \vec \sigma}
= \frac{1}{2}
\left(
\begin{array}{cc}
1-i & -1-i \\
1-i & 1 + i
\end{array}
\right),
\end{equation}
with $\vec \sigma = (\sigma_x,\sigma_y,\sigma_z)$
being a vector of local Pauli operators.
Applying $U$ to all the qubits
gives a new state $| \psi_1 \rangle \equiv U^{\otimes 6} | \psi_3 \rangle$.
With the double application
one gets $| \psi_2 \rangle \equiv U^{\otimes 6} | \psi_1 \rangle$.
The states $| \psi_1 \rangle$ and $| \psi_2 \rangle$
are, up to a global phase which does not contribute to correlations, 
of the same form as $| \psi_3 \rangle$,
but are written in the local bases of $\sigma_x$ and $\sigma_y$
operators, respectively.
Finally, one introduces a mixture
of Greenberger-Horne-Zeilinger correlations
and white noise:
\begin{equation}
\rho = \frac{f}{3} \sum_{k=1}^3
|\psi_k \rangle \langle \psi_k|
+ (1-f) \rho_{\rm noise}.\label{RHO}
\end{equation}
We are interested in six-qubit correlations of this state.
The correlation tensor has $3 \cdot \left({6 \choose 2} + {6 \choose 4} \right) + 3 = 93$ 
nonvanishing six-qubit components.
These are ${6 \choose 2} + {6 \choose 4}$
components with two equal indices 
different than the remaining four equal indices,
e.g., $T_{111122},T_{121121},T_{112222},\ldots$
There are three such sets
which correspond to the three possible
different pairs of indices, i.e., 
$\{1,2\},\{1,3\}$ and $\{2,3\}$ (e.g., 
$T_{111122}, T_{113311},T_{223333},\ldots$).
In the remaining three components
all indices are the same.
The value of every component is given by $\pm f/3$.
Thus, the maximal possible component of the correlation tensor
is equal to $T_{\max} = f/3$.

For certain noise admixture the mixed state (\ref{RHO})
admits a local realistic model 
for correlations obtained in a Bell experiment
with any two local settings.
The sufficient condition
for the existence of such a model
is that the components of the correlation tensor,
maximized over the choice
of all local coordinate systems,
 satisfy \cite{Zukowski}:
\begin{equation}
\max \sum_{i_1,...,i_6 = 1}^2 T_{i_1...i_6}^2 \le 1.
\label{SUFF_STANDARD}
\end{equation}
The state has $32$
components which contribute to this sum.
Thus, the left-hand side equals $\frac{32}{9} f^2$
and the condition is satisfied
for $f \leq \frac{3}{4\sqrt{2}} = 0.53033$.

However, one can still observe non-local behavior of the state
if measurements of more local settings are allowed
even though one add more noise to the state.
First consider a Bell experiment
in which all settings from
arbitrary chosen local planes are measured.
A similar technique to the one described
here (with less general integrations)
leads to the 
necessary condition
for local realistic models which are rotationally invariant 
with respect to the measured correlations \cite{Nagata}:
\begin{equation}
\max \sum_{i_1,...,i_6 = 1}^2 
T_{i_1...i_6}^2 \le \left( \frac{4}{\pi} \right)^6 T_{\max},
\label{2D_NECC}
\end{equation}
where now the maximization is taken
over all possible positions of local measurement planes,
and $T_{\max}$ is computed in
the plane for which the left-hand side is maximal.
For the state under consideration the left-hand side of this condition
is the same as the left-hand side of (\ref{SUFF_STANDARD}),
and one directly finds that the necessary condition (\ref{2D_NECC})
is violated for $f > 0.399422$.
For lower values of $f$
it could be that the specific local realistic model,
proven to exist before,
can be extended to measurements within the plane.

Nevertheless, the new inequality increases the range of $f$
for which the extension is impossible.
This is due to the fact
that the settings over the whole
Bloch sphere are allowed.
For the considered state 
the left-hand side of condition (\ref{3D_NECC})
is the sum of $93$ terms, and thus equals $\frac{93}{9} f^2$,
which gives violation of this condition
whenever $f > 0.36744$.

\section{summary}

In summary,
we derived a generalized $N$-particle Bell inequality 
which involves the entire range of settings for each 
of the local measuring apparatuses. 
The new inequality better reveals the impossibility of a local realistic model
for correlations
in a specific quantum state, i.e., 
a mixture of Greenberger-Horne-Zeilinger states
than many previous inequalities.
We illustrate this by the six-qubit state.
In this case, for a certain noise admixture, 
one can explicitly build
a local realistic model
for the correlations obtained
in a standard Bell experiment
-- the experiment with two local settings
-- independently of the plane which is spanned by the settings.
The inequalities which take
into account the entire range of settings in local planes
disprove the possibility of the model
for a substantially bigger range of noise admixture.
This range can be further enlarged
using inequalities which involve
correlations between observables
from the whole Bloch sphere.

It is very interesting to consider the following.
Could there be
more examples such that this Bell inequality is more stringent?
Could this Bell inequality 
distinguish between different classes of multipartite quantum states? 
What about degree of entanglement for these specific quantum states?

\acknowledgments
The author thanks M. \.Zukowski, W. Laskowski, M. Wie{\'s}niak, 
and T. Paterek for helpfull discussions.
This work has been
supported by Frontier Basic Research Programs at KAIST and K.N. is
supported by the BK21 research professorship.

\end{document}